\documentclass[twocolumn,floatfix,prb,aps,showpacs,superscriptaddress]{revtex4-1} 
\bibpunct{[}{]}{,}{n}{}{}

\usepackage{graphicx,epsf,amsmath,amssymb}
\usepackage{dcolumn}
\usepackage{bm,bbm}
\usepackage{enumerate}
\usepackage[usenames,dvipsnames,svgnames,table]{xcolor}
\usepackage[colorlinks,citecolor=blue,urlcolor=blue]{hyperref}

\usepackage{ulem}  

\begin{document}

\title{Conservation of chirality at a junction between two Weyl semimetals}

\author{S. Tchoumakov} 
\affiliation{Univ. Grenoble Alpes, CNRS, Grenoble INP, Institut N\'eel, 38000 Grenoble, France}

\author{B. Bujnowski}
\affiliation{Donostia International Physics Center (DIPC) - Manuel de Lardizabal 5, E-20018 San Sebasti\'{a}n, Spain}
\affiliation{Univ. Bordeaux, CNRS, LOMA, UMR 5798, F-33405 Talence, France}

\author{J. Noky} 
\affiliation{Max Planck Institute for Chemical Physics of Solids, 01187 Dresden, Germany}

\author{J. Gooth} 
\affiliation{Max Planck Institute for Chemical Physics of Solids, 01187 Dresden, Germany}

\author{A. G. Grushin} 
\affiliation{Univ. Grenoble Alpes, CNRS, Grenoble INP, Institut N\'eel, 38000 Grenoble, France}

\author{J. Cayssol}
\email{jerome.cayssol@u-bordeaux.fr}
\affiliation{Univ. Bordeaux, CNRS, LOMA, UMR 5798, F-33405 Talence, France}

\date{\today}

\begin{abstract}
In Weyl semimetals the location of linear band crossings, the Weyl cones, is not bound to any high symmetry point of the Brillouin zone, unlike the Dirac nodes in graphene. This flexibility is advantageous for valleytronics, where information is encoded in the valleys of the band structure when intervalley scattering is weak. However, if numerous Weyl cones coexist the encoded information can decohere rapidly because of band mixing. Here, we investigate how the helical iso-spin texture  of Weyl cones affects valleytronics in heterojunctions of Weyl materials, and show how the chirality of this iso-spin texture can serve to encode information.
\end{abstract}


\maketitle

An appealing strategy to realize solid-state computations
is to take advantage of the available degrees of freedom displayed by electrons in solids.
Spintronics, for example, is based on manipulating the spin of electrons with magnetic fields to perform switches and memories~\cite{Wolf1488,Cerletti_2005,Smejkal:2018uf,RevModPhys.76.323}. Similarly, valleytronics relies on the several different wavevectors of conducting electrons in a crystal to process and store information~\cite{isberg2013generation,mak2012control,zeng2012valley,li2020room}. For example, the linear band crossings, or Dirac nodes, in graphene can serve as internal degrees of freedom that can be manipulated using light, strain or electric gates~\cite{rycerz2007valley,PhysRevLett.116.016802,gorbachev2014detecting}. Valleytronics can also be envisioned in Weyl semimetals, where the three-dimensional (3D) band crossings, the Weyl nodes, can occur at any point of the Brillouin, unlike their two-dimensional (2D) counterpart~\cite{yesilyurt2019electrically,EREMENTCHOUK20172866}. The flexibility of 3D platforms can be beneficial to valleytronics but also detrimental, as multiple Weyl nodes can overlap and complicate valleytronics~\cite{grassano2018validity}. 

The transmission of electrons at the junction between two Weyl semimetals not only depends on the location of the Weyl nodes, but also on the relative iso-spin textures of the overlapping Fermi surfaces. Indeed, a Weyl node is not only a conical band dispersion that can be located anywhere in the Brillouin zone, but it also carries an helical iso-spin texture ${\bf S}_{\bf k} = \langle \psi_{\bf k} | \hat{\sigma} | \psi_{\bf k} \rangle$ of its eigenstates $|\psi_{\bf k}\rangle$ that winds on the Bloch sphere as a function of the wavevector ${\bf k}$~\cite{liao2020materials}. Since this texture can be different between overlapping cones on either side of a junction, it can affect the electronic transport between two Weyl materials. This helicity-dependent transport, where the momentum dependence of iso-spin ${\bf S}_{\bf k}$ encodes information, is at the intersection between spintronics and valleytronics~\cite{PhysRevB.96.245410,PhysRevApplied.13.054043}.
In particular, the iso-spin texture has a non-vanishing flux over the Fermi surface of a Weyl cone. A Weyl material contains an equal number of Weyl cones with positive and negative flux~\cite{NIELSEN198120,NIELSEN1981173}, that are associated with two chiralities

The multiple Weyl nodes of a Weyl semimetal can be located at various momenta in the Brillouin zone and may have a multitude of iso-spin textures, eventually complicating valleytronics and spintronics because of the many degrees of freedom at hand. In contrast, the chirality of Weyl cones takes only two values, and it is 
thus appealing to track the conservation of chirality of different Weyl nodes across a junction as a means to encode information. However, this poses the challenge of understanding how chirality is transferred taking into account the possible differences in iso-spin textures across the junction. This is the question we investigate in this work.

In the present manuscript, we compute the transport properties of a junction between two Weyl semimetals to show how they depend on the valley, helicity and chirality mismatch. In Sec.~\ref{sec:model} we model the junction between two Weyl semimetals and compute the transmission coefficient of an incoming wavepacket. We first discuss the situation where only two Weyl nodes overlap at the junction and then the situation where many Weyl nodes overlap. In Sec.~\ref{sec:conductance} we compare the amplitude of the conductance for different separations and spin textures of the Weyl nodes. Finally, in Sec.~\ref{sec:discussion} we discuss which materials and configurations are the most suitable to observe valley- and chiral-tronics.

\section{Scattering at the junction between Weyl materials}
\label{sec:model} 

\subsection{Model}

The electronic excitations of a Weyl semimetal, close to a Weyl cone, are described by the Weyl Hamiltonian
\begin{equation}
	\label{eq:model}
    \mathcal{H}(\mathbf{k})= \mathbf{h}(\mathbf{k})\cdot\hat{\boldsymbol{\sigma}} - \epsilon_0 \hat{\mathbbm{1}},
\end{equation}
where $\mathbf{h} = (v_x (k_x - b_x), v_y (k_y - b_y), v_z (k_z - b_z))$ and $\hat{\boldsymbol{\sigma}} = (\hat{\sigma}_x,\hat{\sigma}_y,\hat{\sigma}_z)$ are Pauli matrices associated to an internal degree of freedom that we refer to as iso-spin. The spectrum of Eq.~\eqref{eq:model} is $E_{\sigma}({\bf k}) = -\epsilon_0 + \sigma |{\bf h}({\bf k})|$ which describes a Weyl cone centred at energy $-\epsilon_0$ and wavevector ${\bf b} = (b_x,b_y,b_z)$. The index $\sigma= \pm$ denotes the conduction and valence bands respectively and the linear dispersion is characterized by the velocities ${\bf v} = (v_x,v_y,v_z)$.  The corresponding eigenmodes read
\begin{align}
    \Psi_{\bf k}^{\sigma}( \mathbf{r}) = 
    \frac{1}{\sqrt{2}}\left(
        \begin{array}{c}
            \alpha^{\sigma}({\bf k})\\
            \beta^{\sigma}({\bf k})
        \end{array}
    \right)  \,\, ,
    \label{eq:eigenvector}
\end{align}
where $\phi = {\rm arg}(h_x + ih_y)$ and 
\begin{align}
    &\alpha^{\sigma}({\bf k}) = \left( 1 + \sigma \frac{h_z({\bf k})}{|{\bf h}({\bf k})|} \right)^{1/2},
    \label{eq:alpha}\\
    &\beta^{\sigma}({\bf k}) = \sigma \left( 1 - \sigma \frac{h_z({\bf k})}{|{\bf h}({\bf k})|} \right)^{1/2} e^{i\phi}.
    \label{eq:beta}
\end{align}
In the following we focus on electron dynamics in the conduction band, where ${\sigma} = +$, and drop the mention to the index $\sigma$. The orientation ${\bf S}_{\bf k} = \langle \psi_{\bf k} | \hat{\sigma} | \psi_{\bf k} \rangle$ of the eigenmodes on the Bloch sphere as a function of momentum depends on the sign and amplitude of the velocities $\mathbf{v} = (v_x,v_y,v_z)$. The orientation of ${\bf v}$ is associated with various helical iso-spin textures at the Fermi surface which can be characterized by their chiralities $\chi = {\rm sign}(v_xv_yv_z) = \pm 1$. A Weyl material can have many Weyl points, that come in pairs of opposite chiralities~\cite{NIELSEN198120,NIELSEN1981173}. They can be centred at various momenta and energies, and a Weyl semimetal is said to be chiral if there is a larger density of states of quasiparticles of one chirality over the other. This situation occurs in crystals with low enough symmetries, such as chiral crystals where all mirrors are absent\cite{bradlynScience2016,tangPRL2017,changPRL2017,changNatMat2018,raoNature2019, sanchezPRL2019,takanePRL2019,schroterNatPhys2019}.

\begin{figure}[t]
\includegraphics[width=\columnwidth]{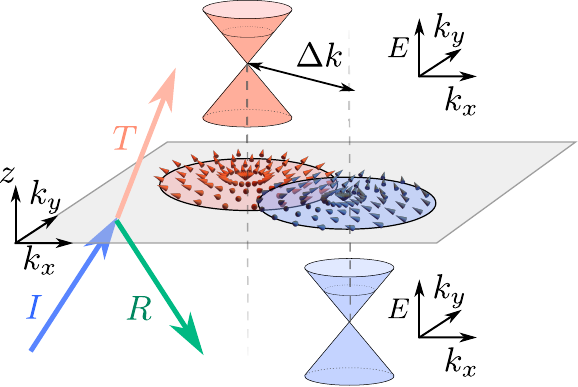}
   \caption{Weyl semimetal junction. At each side of the interface (gray plane), we illustrate the bandstructures of the two Weyl semimetals (red and blue cones) and the overlap of their Fermi seas (red and blue circles), with the associated iso-spin textures. An incoming wavepacket (blue arrow) is either transmitted (red arrow) or reflected (green arrow). Transmission occurs if two conditions are satisfied : (1) that the projection of the Fermi surfaces on the $(k_x,k_y)$ plane overlap, and (2) that eigenspinors are non-orthogonal.}
   \label{fig:sketch}
\end{figure}

In the following we model the scattering of electrons at a junction between two  chiral Weyl materials. We model each chiral Weyl semimetal by considering different energy shifts, $\epsilon_{0}^{\chi}$, on cones of opposite chiralities, $\chi = \pm$, but with a common amplitude for the velocities, $|v_x| = |v_y| = |v_z| = 1$ and allow $\mathbf{v}$ to point in arbitrary directions. The situation with anisotropic velocities can be recovered by rescaling momenta. We first consider the situation where electrons only scatter between two cones, with a single cone on either side of the interface. We then model the situation where the scattering occurs between multiple cones.

\begin{figure*}[htb]
\includegraphics[width=\textwidth]{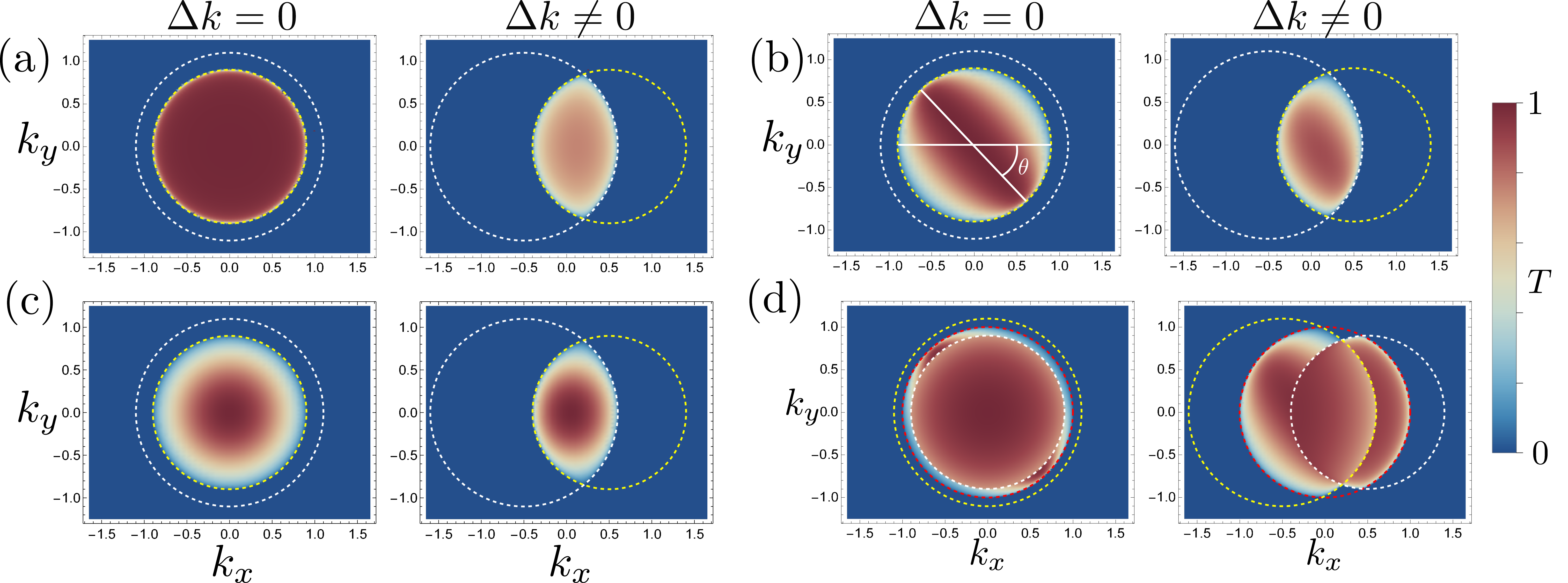}
   \caption{(a,b,c) Transmittance $T$ for each in-plane wavevector ($k_x,k_y$) for the scattering between two Weyl nodes, with energy shifts $\epsilon_{0,L} = 0.9$ and $\epsilon_{0,R} = 1.1$ on the left and right side respectively. The projected Fermi surfaces are depicted with the yellow and white dashed circles. Left columns are for cones not shifted in momentum, $\Delta k = 0$, while this shift is non-zero on the right columns. The figures are obtained for different velocities between the left and right sides of the junction (a) ${\bf v}_{R} = {\bf v}_{L}$, (b) ${\bf v}_{R} = (v_{x,L},v_{y,L},-v_{z,L})$ and (c) ${\bf v}_{R} = (-v_{x,L},-v_{y,L},v_{z,L})$. (d) Transmittance $T$ when scattering from one to two cones, to illustrate the overlap between situations (a) and (b).}
   \label{fig:transmittance}
\end{figure*}

\subsection{Scattering between two Weyl cones}

We consider a sharp interface at $z = 0$ where a Weyl node is shifted in momentum and energy from ${\bf b}_L$ and $\epsilon_{0L}$ for $z < 0$, to ${\bf b}_R$ and $\epsilon_{0R}$ for $z > 0$. This situation is illustrated in Fig.~\ref{fig:sketch}, where blue and red cones describe the electron gas on each side of the interface. These overlapping Weyl nodes can have have different velocities ${\bf v}_{L,R}$, related to different helical spin textures that we depict in Fig.~\ref{fig:sketch} with arrows. 

We suppose electrons scatter elastically, i.e. they conserve their energy $E$. Also, owing to the translation invariance of the interface, at $z = 0$, the components of the wavevectors parallel to the interface ${\bf k}_{\parallel} = (k_x,k_y)$ are conserved. This leads to the following set of equalities
\begin{align}
	\label{eq:snell}
	&k_{z,i} \equiv -k_{z,r} =\\
	&\frac{1}{|v_{z,L}|}\sqrt{ (E - \epsilon_{0L})^2 - v_{x,L}^2(k_x - b_{x,L})^2 - v_{y,L}^2(k_y - b_{y,L})^2 },\nonumber\\
	&k_{z,t} = \nonumber\\
	&\frac{1}{|v_{z,R}|}\sqrt{ (E - \epsilon_{0R})^2 - v_{x,R}^2(k_x - b_{x,R})^2 - v_{y,R}^2(k_y - b_{y,R})^2  },\nonumber
\end{align}
where $k_{z,i}$, $k_{z,r}$ and $k_{z,t}$ are respectively the incoming, reflected and transmitted wavevectors normal to the interface. The conservation of the probability current ${j}_z = v_z \psi^{\dagger} \hat{{\sigma}}_z \psi$ normal to the interface can be expressed as a linear transformation of the wavefunction when crossing the interface, like $\psi_L(z=0) = \hat{g}\psi_R(z=0)$ where $\hat{g}$ is a matrix and $\psi_{L,R}$ are respectively the components of the wavefunction for $z<0$ and $z> 0$ (see Appendix~\ref{app:1to1boundary}).

In the situation where $v_{z,L}/v_{z,R} > 0$, the conservation of current is satisfied by the continuity of the wavefunction $\psi_L(z=0) = \psi_R(z=0)$, so an eigenstate of energy $E$ satisfies
\begin{align}
	\left(
 	\begin{array}{c}
 		\alpha_{L}({\bf k}_{\rm i})  \\
 		\beta_{L}({\bf k}_{\rm i}) 
 	\end{array}
 	\right) 
 	+
 	r \left(
 	\begin{array}{c}
 		\alpha_{L}({\bf k}_{\rm r}) \\
 		\beta_{L}({\bf k}_{\rm r}) 
 	\end{array}
 	\right)
 	= 
 	t \hat{M} \left(
 	\begin{array}{c}
 		\alpha_{R}({\bf k}_{\rm t }) \\
 		\beta_{R}({\bf k}_{\rm t }) 
 	\end{array}
 	\right).
 	\label{eq:linj1}
\end{align}
where $\hat{M} = \hat{\mathbbm{1}}$. The $L,R$ subscript of $\alpha({\bf k})$ and $\beta({\bf k})$ remind us that Eqs.~(\ref{eq:alpha},\ref{eq:beta}) should be evaluated for the parameters on the $z < 0$ and the $z > 0$ half-space respectively. 
We then find that
\begin{equation}
	r = \frac{\beta_{L}({\bf k}_{\rm i})\alpha_{R}({\bf k}_{\rm t }) - \alpha_{L}({\bf k}_{\rm i})\beta_{R}({\bf k}_{\rm t }) }{\alpha_{L}({\bf k}_{\rm r})\beta_{R}({\bf k}_{\rm t }) - \beta_{L}({\bf k}_{\rm r})\alpha_{R}({\bf k}_{\rm t }) } \, .
\end{equation}
From this expression we compute the reflectance $R = |r|^2$, and the transmittance $T =1-R$. In  Fig.~\ref{fig:transmittance}(a), we plot $T$ as a function of  ($k_x,k_y$) for two different separations, $\Delta k = b_{x,R} - b_{x,L}$, with $b_{y,L/R} = b_{z,L/R} = 0$, between the two Weyl nodes on each side of the interface with ${\bf v}_L = {\bf v}_R$.
Since electrons only scatter through the interface when there are available states at the same energy and in-plane momenta, the transmission probability is non-zero only where the two Fermi surfaces, depicted by dashed lines, overlap~\cite{EREMENTCHOUK20172866}. Also, we observe that when the two Fermi surfaces are shifted ($\Delta k \neq 0$) the transmittance is smaller than for zero shift ($\Delta k = 0$), because of the imperfect overlap between spinors on each side of the interface. Indeed, as we increase $\Delta k$, the iso-spin textures on each side of the junction progressively point in opposite directions along the $x-$axis, suppressing $T$.

In the situation where $v_{z,L}/v_{z,R} < 0$, the continuity equation~\eqref{eq:linj1} with $\hat{M} = \hat{\mathbbm{1}}$ does not apply since it breaks current conservation. In appendix~\ref{app:1to1boundary}, it is shown that the correct boundary condition requires to invert the $z$ component of the spinor across the interface. This is related to the lack of a solution for a hyperbolic equation governing the conservation of current, since there is no Lorentz boost to map the two Weyl equations. In this case, the matrix $\hat{M}$ on the right-hand side of Eq.~\eqref{eq:linj1} is instead 
\begin{align}
	\hat{M} = \cos(\theta)\hat{\sigma}_x + \sin(\theta)\hat{\sigma}_y,
 	\label{eq:linj2}
\end{align}
where $\theta \in [0,2\pi)$ is a parameter associated with the junction. In our model the parameter $\theta$ is arbitrary and in a realistic heterojunction it is set by the microscopic coupling between the two Weyl materials.

As above we compute the reflection coefficient $r$, reflectance $R = |r|^2$ and transmittance $T = 1 - R$. In Fig.~\ref{fig:transmittance}(b), we plot $T$ as a function of ($k_x,k_y$) for different values of $\Delta k$, keeping $v_x = v_y = 1$ across the interface but with $v_{z,L} = -v_{z,R}$. We see that the transmittance is cigar-shaped, with a principal axis in the direction of $\theta$. This asymmetry is a consequence of the rotation in Eq.~\eqref{eq:linj2}, because it not only inverts the $z$-component of iso-spin but also the component along ${\bf s}_{\perp} = -\sin(\theta) {\bf e}_x + \cos(\theta) {\bf e}_y$. Thus, for an in-plane momentum $(k_x,k_y)$ along ${\bf s}_{\perp}$, the eigenspinors on each side of the interface overlap poorly and lead to a smaller transmission even if the Fermi surfaces match, \textit{i.e.} even if $\Delta k = 0$.

In this section we have considered two situations depending on whether $v_z$ changes sign or not, keeping $v_x$ and $v_y$ the same across the interface.
These two situations demonstrate the importance of the spin texture, set by $\bf{v}$, when electrons scatter across the junction formed by two Weyl semimetals. In the following we discuss the more general situation where all velocities can change direction and how transmittance depends on the chiralities of Weyl cones.

\subsection{Role of chirality on scattering}
\label{sec:chiralities}

As discussed in the introduction, a central feature of Weyl cones in Weyl semimetals is there chirality, $\chi = {\rm sign}(v_xv_yv_z)$. The chirality of a Weyl node only takes two values, $+1$ and $-1$. It is thus a simpler description of the iso-spin texture of Weyl nodes compared to a continuous set of iso-spin configurations described by ${\bf v}$. In this respect, the two situations of the previous section correspond to the transmission between cones with same chiralities, when $v_{z,L}/v_{z,R} > 0$ in Fig.~\ref{fig:transmittance}(a), or with opposite chiralities, when $v_{z,L}/v_{z,R} < 0$ in Fig.~\ref{fig:transmittance}(b).

The junction between Weyl cones with same chiralities happens when ${\bf v}_{L}$ and ${\bf v}_R$ are related by a rotation. This leads to the transmission function in Fig.~\ref{fig:transmittance}(a) when ${\bf v}_{L} = {\bf v}_R$, or to that in Fig.~\ref{fig:transmittance}(c) when ${\bf v}_{R} = ( -v_{x,L}, -v_{y,L}, v_{z,L})$, \textit{i.e.} for a $\pi$-rotation around the $z-$axis. In general, the transmission function is rotation-invariant but its radial distribution depends on the angle between ${\bf v}_{L}$ and ${\bf v}_R$. For example, in Fig.~\ref{fig:transmittance}(c) at $\Delta k = 0$, the transmission drops on the borders of the Fermi surface because there the spinors of the two Weyl cones point in opposite directions.

The junction between Weyl cones with opposite chiralities happens when ${\bf v}_{L}$ and ${\bf v}_R$ are related by combining an inversion with a rotation. This leads to a cigar-shaped transmission function, similar to Fig.~\ref{fig:transmittance}(b), with an angle $\theta$ that depends on the relative orientation between ${\bf v}_{L}$ and ${\bf v}_R$. In general, the transmission function breaks rotation symmetry in the $(k_x,k_y)$ plane, even for $\Delta k = 0$. This absence of rotation symmetry is related to the mismatch of the iso-spin textures on either side of the interface along a single axis, in the direction ${\bf s}_{\perp} = -\sin(\theta) {\bf e}_x + \cos(\theta) {\bf e}_y$.

Through this discussion, we find that there is a qualitative difference between the transmission of electrons between Weyl cones with the same or opposite chiralities. On one hand, if chiralities are the same then $T(k_x,k_y)$ is rotational invariant and only its radial distribution can change. On the other hand, if chiralities are opposite then $T(k_x,k_y)$ breaks rotational invariance and its radial distribution is always the same, up to a rotation in the ($k_x,k_y$) plane.

\subsection{Scattering between multiple Weyl cones}
\label{sec:multiple}

The previous model concerns the scattering between only two Weyl cones. Since Weyl materials usually host many Weyl cones, the previous discussions can break down once multiple Weyl nodes contribute to scattering. The transmission depends on helicity and may not split evenly between each scattering channel. In this section we consider an heterojunction where the electron gas scatters from a single Weyl node, at $z < 0$, to two Weyl nodes, at $z > 0$. 

We model this interface assuming that the in-plane momentum and the energy are conserved, and apply the conservation relations in Eq.~\eqref{eq:snell} independently for each Weyl cone. The conservation of the probability current is discussed in Appendix~\ref{app:1tomanyweyls} and it involves a free parameter, $m_{12} = T_1/T_2$, related to the relative contribution of the two Weyl nodes at $z > 0$ to scattering. Like the parameter $\theta$ in Eq.~\eqref{eq:linj2}, $m_{12}$ depends on the precise description of the interface, such as the orbitals involved and their overlap. In the following we assume that $m_{12} = 1$, corresponding to an equal transmission towards any of the two cones at $z > 0$. The reflectance and transmittance of the incoming wavepacket are obtained by solving linear equations similar to Eq.~\eqref{eq:linj1}. 
The general expression for the transmittance is provided in Appendix~\ref{app:1tomanyweyls}.

We observe that the transmittance for scattering from one Weyl node at $z < 0$ to two Weyl nodes at $z > 0$ is not the average transmittance for scattering to each Weyl node individually. For example, in Fig.~\ref{fig:transmittance}(d) we show the transmittance for ${\bf v}_{R}^{(1)} = {\bf v}_{L}$ and ${\bf v}_{R}^{(2)} = (v_{x,L}, v_{y,L}, -v_{z,L})$, which is similar to the superposition of cases reported in Fig.~\ref{fig:transmittance}(a) and (b). Thus, our observation in Sec.~\ref{sec:chiralities}, that the qualitative behaviour of transmittance between two cones with same and opposite chiralities are different, is not satisfied when scattering between multiple Weyl cones. The scattering between multiple Weyl cones does not conserve chirality and strongly depends on the interface-dependent parameter $m_{12}$, so this configuration should be avoided when using chirality as a mean to carry information.

\section{Conductance}
\label{sec:conductance}

\begin{figure}[t!]
    \centering
    \includegraphics[width=\columnwidth]{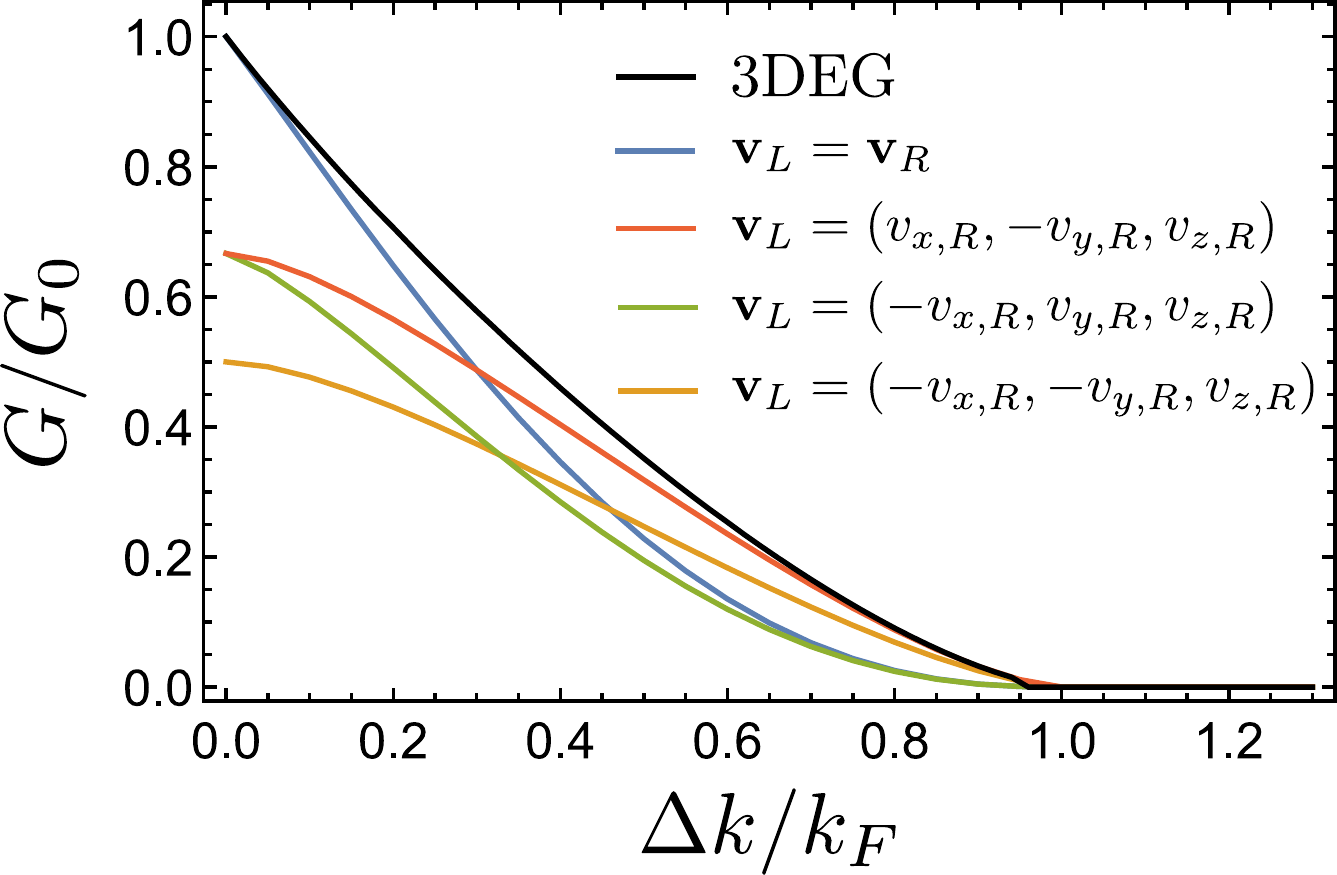}
    \caption{Behaviour of the conductance across a junction of two Weyl material as a function of the shift $\Delta k = b_{x,R} - b_{x,L}$ between the band structures along the $x-$axis, with $b_{y,L/R} = b_{z,L/R} = 0$. The different curves are for each relative iso-spin textures in Fig.~\ref{fig:transmittance}(a-c), that we label with their velocities ${\bf v}_{L,R}$. In black, we show the transport at a junction between two non-relativistic bandstructures (3DEG), with Hamiltonian $\hat{H}_{NR} = \hbar^2 (k-\Delta k)^2/(2m)$.}
    \label{fig:conductancesdk}
\end{figure}

The electric current $I$ through the junction between two Weyl materials can be computed within the Landauer B\"{u}ttiker formalism as
\begin{align}
	I = \frac{e}{h}\int dE \sum_{k_x,k_y} T(E,k_x,k_y)\left(f_L(E)-f_R(E)\right)~,
	\label{eq:lanbut}
\end{align}
where $T(E,k_x,k_y)$ is one of the transmittances depicted in Fig.~\ref{fig:transmittance}. The function $f_\alpha(E)$ is the Fermi-Dirac distribution in lead $\alpha=L,R$, with chemical potential $\mu_\alpha$.
At zero temperature $T=0$ and for a small bias voltage $V$ at the junction ($\mu_L=\mu_R + eV$), the differential conductance reads
\begin{flalign}
	G \equiv \frac{\partial I}{\partial V}=\frac{e^2 W^2}{h} \iint \frac{dk_x dk_y}{(2\pi)^2} \, \, T(E = 0,k_x,k_y)    \, ,
\end{flalign}
and depends only on the transmittance at the Fermi surface. The area of the junction is $W^2$ and it is assumed that $W$ is larger than the Fermi wavelength so that the discrete sum over $(k_x,k_y)$, defined in integer multiples of $2\pi/W$, can be replaced by an integral. The range of integration is set by the region where the Fermi surfaces on each side overlap (see Fig.~\ref{fig:transmittance}), \textit{i.e.} where $k_z$ in Eq.~\ref{eq:snell} is real.

In Fig.~\ref{fig:conductancesdk} we show the conductance of our model as a function of the Weyl node shift $\Delta k$ across the interface. The conductance is given in units of 
\begin{align}
	G_0=\pi \frac{e^2 }{h}  \frac{W^2 \mu_0^2 }{(2\pi\hbar v_F)^2} \,,
\end{align}
which is the conductance of a single Weyl node at chemical potential $\mu_0$, without the interface. The chemical potential $\mu_0$ is our reference for chemical potentials, which we define to be the geometric average of all energy shifts $\epsilon_0$ across the interface. The scattering between cones vanishes when they are too far apart in momentum space. This property was discussed in Ref.~\cite{EREMENTCHOUK20172866} as a way to transmit current between overlapping Weyl cones, independently of their iso-spin texture, and it illustrates valleytronics in Weyl semimetals. 

Also in Fig.~\ref{fig:conductancesdk}, we compare the different configurations whose single-channel transmittances were illustrated in Figs.~\ref{fig:transmittance}(a-c), and show that the relative iso-spin textures strongly affect transport. The conductance between two Weyl semimetals appears to be always smaller than the conductance between two non-relativistic metals with the same carrier densities, as a consequence of the non-colinear iso-spin textures.

\begin{figure}[t!]
    \centering
    \includegraphics[width=\columnwidth]{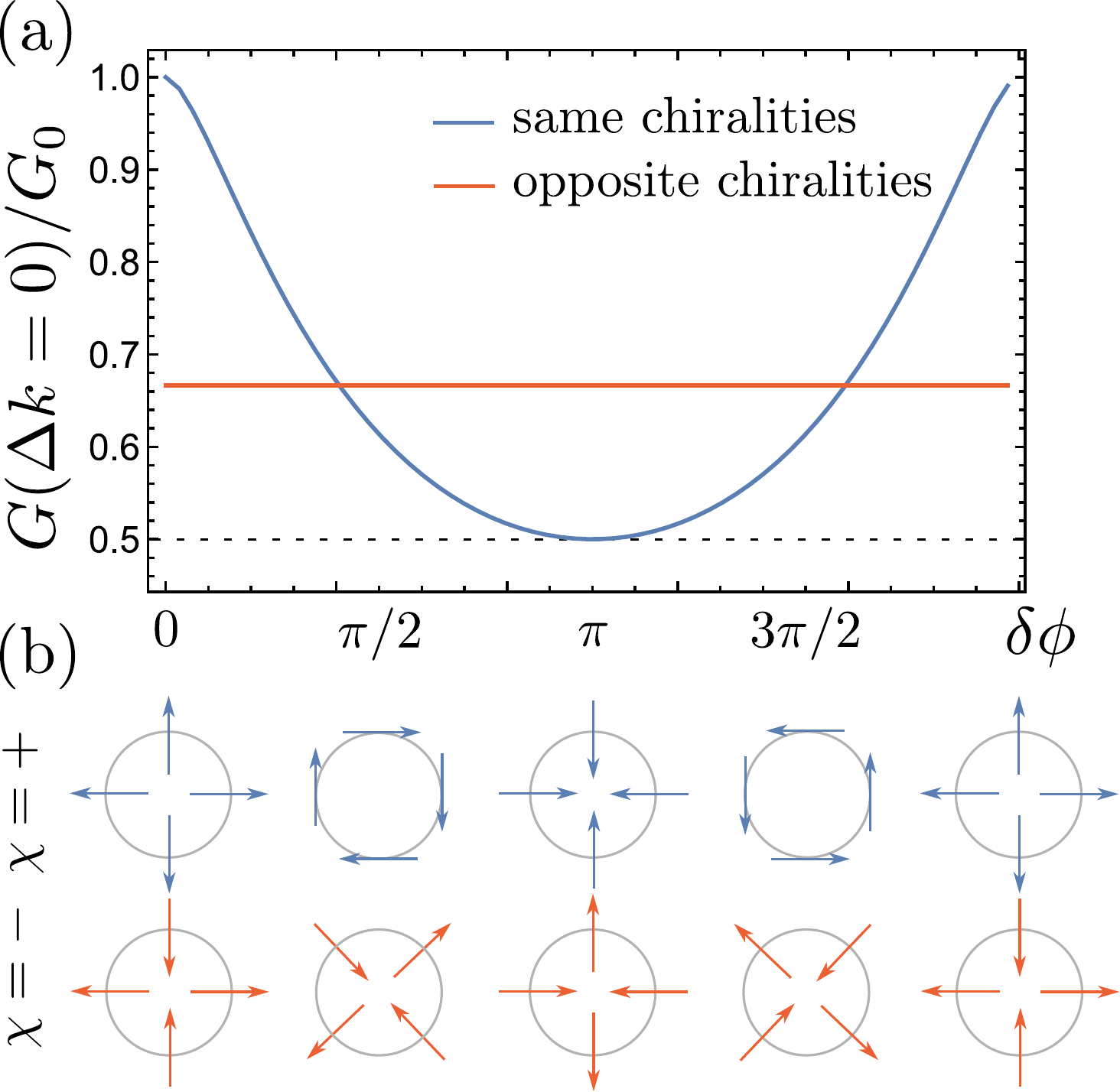}
    \caption{(a) Conductance across a junction between two Weyl materials, with $\Delta k = 0$, as a function of the phase shift $\delta \phi$ between the direction of momentum and iso-spin in the $xy$ plane. The conductance between cones with same chirality strongly depends on their relative angle. (b) We illustrate the change in the iso-spin texture as a function of the phase shift $\delta \phi$ of the Weyl material at $z > 0$ and for two opposite chiralities ($\chi = \pm 1$). The horizontal axis is the same in (a) and (b).}
    \label{fig:conductancestheta}
\end{figure}

In Fig.~\ref{fig:conductancestheta}(a) we show the conductance of our model for $\Delta k = 0$, when Weyl cones exactly overlap in momentum space, as a function of the phase shift $\delta \phi$ between momentum and the iso-spin texture in the $xy$ plane. This phase shift is introduced in Eq.~\eqref{eq:beta} by replacing the phase $\phi = \arg(h_x+ih_y)$ with 
\begin{align}
    \phi \rightarrow \phi + \delta \phi.
\end{align}
The transmission between cones with opposite chiralities (in red) does not depend on the phase-shift $\delta \phi$ while it does for scattering between cones with same chiralities (in blue). This can be intuitively understood from Fig.~\ref{fig:conductancestheta}(b) that shows the change in the helical spin texture on the projected Fermi sea as a function of $\delta \phi$. For an interface between cones of the same chirality the relative in-plane iso-spin texture can either point inward or outward as a consequence of this phase-shift, leading to important changes in the iso-spin overlap like illustrated in Fig.~\ref{fig:transmittance}(a,c). The transmittance between Weyl nodes with same chirality is minimal for $\delta \phi = \pi$, when the in-plane iso-spin textures point in opposite directions on either side of the interface. For an interface between cones of opposite chiralities the relative in-plane iso-spin texture always points outward in one direction while inward in the other. In this case, the overall overlap between helical iso-spin textures does not change, \textit{i.e.} Fig.~\ref{fig:transmittance}(b) is solely rotated, and the conductance is independent on $\delta \phi$. 

This way, the relative contribution of transport between cones that change or conserve chirality fluctuates. In particular, one can manipulate the phase-shift $\delta \phi$ geometrically for anisotropic iso-spin textures, \textit{i.e.} when there are different signs between two of the components of ${\bf v}_L$ or ${\bf v}_R$.

\section{Transport of chirality in materials}

\label{sec:discussion}

\begin{figure*}[t]
    \centering
    \includegraphics[width=\textwidth]{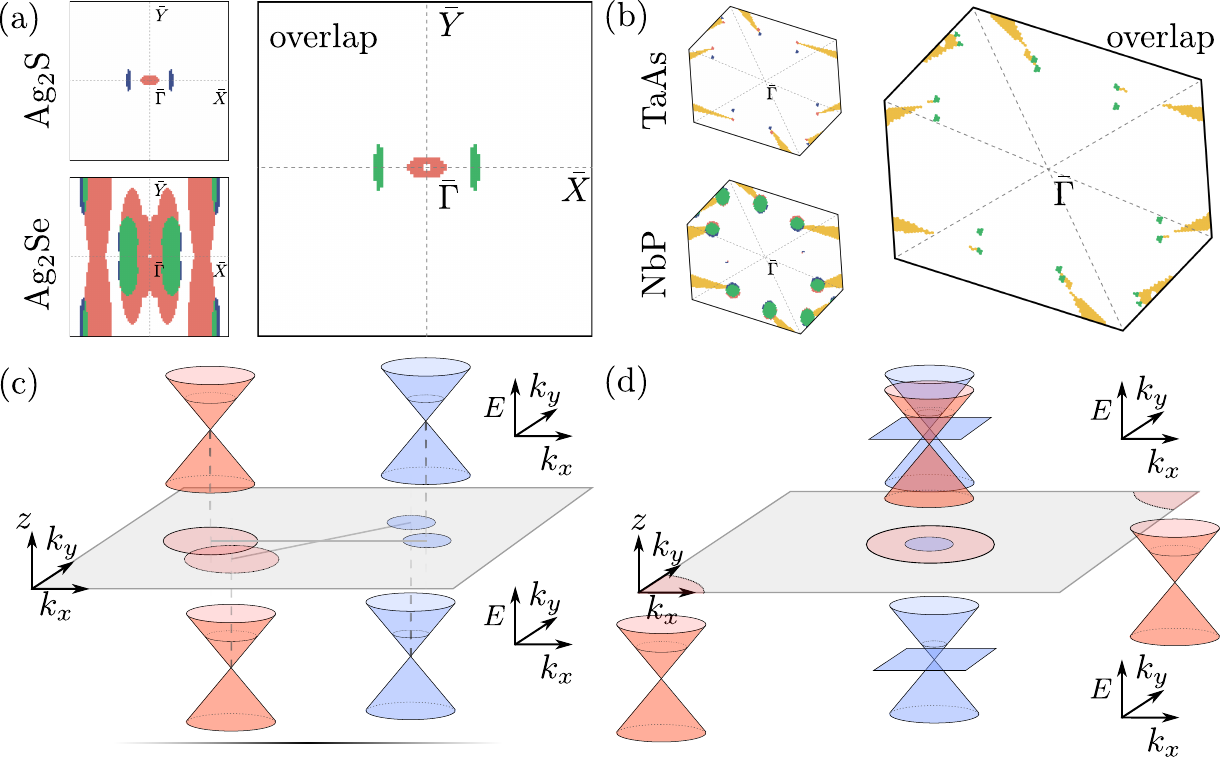}
   \caption{(a,b) Transport at the interface of (a) Ag$_2$S and Ag$_2$Se and (b) TaAs and NbP. We draw the Fermi surface of each material with a color that indicates if the chirality is positive (blue), negative (red) or an overlap of both (green). The trivial Fermi pockets are in orange. (c) In a chiral Weyl semimetal, the interface between two orientations of the Weyl semimetal can help transmit only one chirality, which corresponds to cones in red on in this figure. (d) In CoSi, the double Weyl cones (red) are at the border of the Brillouin zone and their location strongly depends on the orientation of the interface.}
   \label{fig:discussion}
\end{figure*}

Our findings may apply to describe junctions of various semimetals such as the transition metal monopnictides TaAs~\cite{PhysRevX.5.031013,grassano2018validity,ramshaw2018quantum}, TaP~\cite{Xue1501092,grassano2018validity}, NbAs~\cite{xu2015discovery,grassano2018validity} and NbP~\cite{PhysRevB.93.161112,grassano2018validity}, or the silver chalcogenides Ag$_2$S~\cite{PhysRevB.100.205117} and Ag$_2$Se~\cite{PhysRevB.96.165148} which have been studied in the development of memristive devices~\cite{C7TA04949H,lee2017forming}. 
These materials show multiple Weyl nodes with rather small carrier densities, $n \approx 10^{17}-10^{19}$~cm$^{-3}$ corresponding to $k_F \approx 0.1-1$~nm$^{-1}$, and located away from the $\Gamma$ point. 
For example, the bandstructure of TaAs contains 24 Weyl nodes located at about $\delta k \approx 10$~nm$^{-1}$ from the $\Gamma$ point.

A particularly clean band structure can be found in chiral semimetals~\cite{bradlynScience2016,tangPRL2017,changPRL2017,changNatMat2018,raoNature2019, sanchezPRL2019,takanePRL2019,schroterNatPhys2019}, specifically in the materials CoSi and RhSi, in space group 198. The band structure in this space group features a topologically protected threeband crossing at the $\Gamma$ point and a doubly degenerate Weyl crossing at the Brillouin zone corner. CoSi and RhSi are especially favourable, as their topological band crossings are close to the Fermi energy, with relatively small or no trivial pockets at this energy. The node separation $\delta k$ is maximum, as it is half of the Brillouin zone ($\delta k=\pi/a$). 
The transport lifetime is shorter in RhSi ($\approx13$ fs)~\cite{Ni:2020uua} than in CoSi ($47$ fs)~\cite{Xu27104,ni2021giant}. For the latter the carrier density is around $2.2 \times 10^{20}$ cm$^{-1}$ corresponding to $k_F = 1.9~$nm$^{-1}$~\cite{Xu27104,ni2021giant}, and the lattice constant is $a=0.4485$ nm, which results in $\delta k = 7$ nm$^{-1}$. An interface between the [001] and [111] surfaces of either CoSi or RhSi is sketched in Fig.~\ref{fig:discussion}(d). In the [001] direction the threefold and double Weyl node project to the center and corner (at $(k_x,k_y) = (\pi/a,\pi/a)$) of the surface Brillouin zone. In contrast, on the [111] surface the threefold and double Weyl fall on top of each other at the center of the Brillouin zone (at $(k_x,k_y) = (\pi/a,\pi/a)$). The position of the double Weyl nodes in the ($k_x,k_y$) plane thus shifts by $\Delta k = \sqrt{2}\pi/a = 9.9~$nm$^{-1}$ between the two orientations.

The junction between two Weyl semimetals in Sec.~\ref{sec:model} can describe the interface between two different Weyl materials or between two orientations of the same Weyl material. We consider these two situations in the following and also discuss the conservation of chirality at the interface with a normal metal.

\subsection{Interface between different Weyl materials}

In Fig.~\ref{fig:discussion} (a) and (b) we show the projection of the Fermi surface in the $[001]$ direction of four Weyl semimetals, respectively Ag$_2$S and Ag$_2$Se and, TaAs and NbP from our ab-initio calculations (see Appendix~\ref{app:abinitio}). We consider these interfaces because of their small lattice mismatch, that should not lead to strong lattice defects at the junction. The Fermi surfaces are colored according to the chirality of the underlying quasiparticles: red and blue for Weyl cones with positive and negative chiralities, green for the superposition of Weyl cones with opposite chiralities and, orange for trivial pockets where chirality is not defined.

The overlap of these Fermi surfaces shows that in monopnictides the chirality of quasiparticles mixes at the interface and that for silver chalcogenides only quasi-particles with positive chirality propagate close to the $\Gamma$ point. The interface between TaAs and NbP also demonstrates that the contribution from trivial pockets is usually non-negligible and is detrimental to well defined chiral-tronics~\cite{grassano2018validity,ramshaw2018quantum}. Our ab-initio calculations show that there are no contribution from trivial pockets in Ag$_2$S and Ag$_2$Se. Thus the interface between these two materials is a good candidate to observe transport with a well-defined chirality of Weyl quasiparticles.

\subsection{Interface between different material orientations}
\label{sec:geometric}
The transmission of chirality between two Weyl semimetals depends on two geometrical aspects, related to the location of Weyl nodes in the Brillouin zone and to their relative iso-spin textures.

A junction can be made of a single Weyl material but with different growth direction or magnetic field orientation on either side of the junction~\cite{liao2020materials}. This can lead to the overlap of Fermi surfaces of a certain chirality but not the other. We illustrate this possibility in Fig.~\ref{fig:discussion}(c). The transmission of a Weyl cone at an interface of angle $\alpha$ between two orientations of a Weyl material, occurs if $\Delta k = 2\delta k \tan(\alpha/2) < k_F$, {\it i.e.} $|\tan(\alpha/2)| < k_F/(2\delta k)$. This critical angle depends on the carrier density of the Weyl cone, through the Fermi wavevector $k_F$. Therefore, selective transmission of chirality can happen in chiral Weyl semimetals since cones of opposite chiralities may have different $k_F$ (see Fig.~\ref{fig:discussion}(c)). The critical angle also depends on the distance $\delta k$ of the Weyl cone to the $\bar{\Gamma}$-point. Therefore, selective transmission of chirality can happen when Weyl nodes of a given chirality are far from the center of the Brillouin zone. For example, this should occur at the interface between the [111] and [100] orientations of CoSi where $\Delta k = 9.9~$nm$^{-1} > 1.9~$nm$^{-1} = k_F$~(see Fig.~\ref{fig:discussion}(d)).

The relative iso-spin textures of the Weyl cones also affects the transport of chirality. For example, at the interface between TaAS and NbP in Fig.~\ref{fig:discussion}(b) we find that Weyl cones with opposite chiralities can overlap and lead to the non-conservation of chirality at the junction. This non-conservation of chirality is independent of the iso-spin textures for overlapping Fermi seas, \textit{i.e.} for $\Delta k = 0$. On the contrary, the contribution to transport between Weyl cones with same chiralities strongly depends on their relative iso-spin textures. In the case of anisotropic iso-spin textures, when velocities are not all of the same sign, this is something that can even be tuned geometrically. From Fig.~\ref{fig:conductancestheta}(b), we see that the contribution to transport that conserves chirality can be tuned from $3/2$ (at $\delta \phi = 0$) to $3/4$ (at $\delta \phi = \pi$) the contribution from channels that change chirality.

\subsection{Interface between a Weyl material and a metal}
\label{sec:normalweyl}

The interface between a normal metal, with a quadratic band dispersion, and a Weyl semimetal should unavoidably appear in transport experiments where metallic leads are used to probe the Weyl semimetals samples. Besides the fundamental issue of understanding how chirality is transmitted to a non-chiral medium, the analysis of the metal to Weyl semimetal interface is crucial to evaluate the contact resistance in transport measurements~\cite{PhysRevLett.102.026807,PhysRevB.78.121402,PhysRevB.79.075428}.

In Appendix~\ref{app:3deg} we model the junction of a metal with a Weyl semi-metal, where we account for the transition from a scalar to a spinorial wavefunction at the interface. The current operators on either side of the interface are also drastically different, being momentum dependent in the metal and iso-spin dependent in the Weyl semimetal. We show that the transmitted current does not depend on the iso-spin texture of the Weyl semimetal, but only on the overlap of its Fermi surface with that of the normal metal. We thus expect that transport across the interface of a metal with a Weyl semimetal is independent on chirality. Also, for $\Delta k = 0$, the conductance of a junction between a Weyl semimetal and a normal metal is smaller than between two normal metals, as a consequence of the imperfect overlap between the iso-spin texture in the Weyl semimetal with that in the normal metal.

\medskip 

\section{Conclusion}
\label{sec:conclusion}

We have discussed how and when chirality, a feature of Weyl quasiparticles, can serve as a well-defined quantity in transport at a junction between two Weyl semimetal. In particular we have discussed how a junction can polarize the current to a single chirality and how chirality is conserved when cones of opposite chiralities overlap.
The polarization of current to a single chirality mostly occurs for materials with few and well separated cones~\cite{yesilyurt2019electrically,EREMENTCHOUK20172866}. We show that this can occur at the junction between the Weyl semimetals Ag$_2$S and Ag$_2$Se along their [001] interface or at the interface between [111] and [100] orientations of CoSi. 
In general, it is difficult to transmit only one chirality with high efficiency because cones are too close to each other or that there are too many of them, a problem we illustrate with the interface between TaAs and NbP. 
We show that transport between overlapping Weyl cones then strongly depends on their respective iso-spin textures. For Weyl nodes of opposite chiralities there is always a non-zero contribution to transport, implying a non-conservation of chirality. However, for Weyl nodes with the same chiralities, the conductance strongly depends on the two iso-spin textures.

\bigskip

{\it Acknowledgements.-} We thank B. Gotsman, H. Schmid and A. Molinari for discussions about experimental details. 
A. G. G and S. T acknowledge financial support from the European Union Horizon 2020 research and innovation program under grant agreement No829044 (SCHINES).
A. G. G. is also supported by the ANR under the grant ANR-18-CE30-0001-01 (TOPODRIVE).  J. C acknowledges support from the Quantum Matter Network at Bordeaux University under project TaQuaMaUC.

\bibliographystyle{apsrev4-1}
\bibliography{bibliography}

\appendix
\begin{widetext}
\section{Boundary conditions}

We derive the boundary conditions of the heterojunction from the conservation of the probability current. This approach has been extensively used to describe the scattering of Dirac electrons with vacuum~\cite{berrymondragon,enaldiev2015boundary,volkov2016surface,PhysRevB.95.081302}, for example in the theoretical modelling of Fermi arcs in Weyl semimetals~\cite{PhysRevB.95.081302}. The minimal constraint to model the boundary with vacuum is that the time evolution be unitary, \textit{i.e.} that the average energy $E = \langle \psi | \hat{H} | \psi \rangle$ of any state $\psi$ is real. This implies
\begin{align}
    E - E^* = - i \oint_{\partial \mathcal V} d{\bf S}\cdot {\bf j}({\bf r}) = 0,
\end{align}
where $\partial \mathcal{V}$ is the boundary of the Weyl medium with vacuum and ${\bf j} = \langle \psi| {\bf v}\hat{\boldsymbol{\sigma}}|\psi\rangle$ is the probability current of a single Weyl cone. In the case of an interface between two materials, this expression is instead
\begin{align}
    E - E^* = - i \sum_{n = L,R}\oint_{\partial \mathcal V} d{\bf S}_n \cdot {\bf j}_n({\bf r}) = 0,
\end{align}
where the index $n$ refers to each medium. In the main text we use $n = L$ for the medium at $z < 0$ and $n = R$ for the medium at $z > 0$. A solution to this equation is ${\bf j}_L = {\bf j}_R$, a quadratic equation between the component $\psi_1$ and $\psi_2$ of the wavefunction on each side of the interface
\begin{align}
    \langle \psi_L | v_{z,L} \hat{\sigma}_z | \psi_L \rangle = \langle \psi_R | v_{z,R} \hat{\sigma}_z | \psi_R \rangle 
    \label{app:eq:current101}
\end{align}
where we suppose that the interface is normal to $z$. In the following we discuss the linearization of this boundary condition, similar to the approach in~\cite{PhysRevB.95.081302}, for the scattering from one Weyl cone to another in 1., the scattering from one Weyl cone to $N$ others in 2. and the scattering of a normal electron gas with a Weyl cone in 3..

\subsection{Scattering between two Weyl cones}
\label{app:1to1boundary}

In the case we scatter an electron from one Weyl cone to another, we can relate the wavefunction on each side of the interface: $\psi_{L}$ for $z < 0$ and $\psi_{R}$ for $z > 0$, by a linear transformation
\begin{align}
    \psi_{L} = \hat{g} \cdot \psi_{R}.
\end{align}
This equation can be interpreted as a change in the reference frame when electrons scatter from one Weyl cone to another. This equation and Eq.~\eqref{app:eq:current101} imply that the matrix $\hat{g}$ satisfies
\begin{align}
    \hat{g}^{\dagger}\hat{\sigma}_z \hat{g} = r \hat{\sigma}_z
\end{align}
where $r = {v_{z,R}}/{v_{z,L}}$, and a solution to this is
\begin{align}
    \hat{g} = a_1 \hat{\mathbbm{1}} + a_2\left( b_1 \hat{\sigma}_x + b_2 \hat{\sigma_y} \right)
\end{align}
with $a_1,a_2,b_1,b_2 \in \mathbbm{R}$, $a_1^2 - a_2^2 = r$ and $b_1^2 + b_2^2 = 1$. We see that there are many possible solutions, determined by surface-related parameters. The parameters $(a_1,a_2)$ are solutions to a hyperbola which can be associated to Lorentz boosts in special relativity, while $(b_1,b_2)$ are on the circle which is associated to rotations. Note that in this work we suppose that the matrix $\hat{g}$ is unitary, a non-unitary transformation would imply some gain or loss of probability density through the boundary.

This way we see that we can distinguish two regimes depending on the sign of $r = {v_{z,R}}/{v_{z,L}}$:
\begin{itemize}
    \item in the situation where $r > 0$, \textit{i.e.} when the normal component of the spin helicity does not change sign, we can always use the solution
    \begin{align}
        \hat{g}_{1} = \sqrt{r} \hat{\mathbbm{1}},
    \end{align}
    \item in the situation where $r < 0$, \textit{i.e.} when the normal component of the spin helicity changes sign, the previous solution is not valid. Instead, we can use the solutions
    \begin{align}
        \hat{g}_{2} = \sqrt{|r|}( b_1 \hat{\sigma}_x + b_2 \hat{\sigma_y} ),
    \end{align}
    with $b_1^2 + b_2^2 = 1$. In the main text we write $(b_1,b_2) = (\cos(\theta),\sin(\theta))$.
\end{itemize}

\subsection{Scattering from one to many Weyl cones}
\label{app:1tomanyweyls}

In the case we scatter an electron from one Weyl cone to $N$ others, we again relate the wavefunction on each side of the interface: $\phi_{L}$ for $z < 0$ and $\phi_{i,R}$ for $z > 0$, where $i \in [1,N]$ indexes each Weyl cone, by an ensemble of linear transformations
\begin{align}
    \psi_{i,R} = \hat{g}_i \psi_{L},
\end{align}
where we suppose $\hat{g}_i$ are invertible. This implies that all the $\phi_i$ are related to each other by a relation of the form
\begin{align}
    \psi_{j,R} = \hat{M}_{ji} \psi_{i,R},
\end{align}
with $\hat{M}_{ji} = \hat{g}_j\hat{g}_{i}^{-1} = m_{ji} \hat{U}_{ji}$ with $m_{ji}$ a scalar, that represents the transmission ratio between $i$ and $j$ channels, and $\hat{U}_{ji}$ a unitary transformation. In the case the transmitted states are eigensolutions of the original Hamiltonian with energy $E$, we find that 
\begin{align}
    \hat{U}_{ij} = \frac{\hat{\mathbbm{1}} + \frac{ \hat{H}_{j}\hat{H}_i } {(E-\epsilon_{0,j})(E-\epsilon_{0,i})} }{\left( {\rm Tr}\left( \hat{\mathbbm{1}} + \frac{ \hat{H}_{j}\hat{H}_i } {(E-\epsilon_{0,j})(E-\epsilon_{0,i})} \right) \right)^{1/2}},
    \label{app:eq:unitary}
\end{align}
where $\hat{H}_{i}$, $\hat{H}_{j}$ are the Hamiltonians of $\phi_{i,R}$ and $\phi_{j,R}$ respectively. These Hamiltonians satisfy Eq.~\eqref{eq:model} with different velocities ${\bf v}_i$, momentum shifts ${\bf b}_i$ and energy shift $\epsilon_{0,i}$.

We insert these relations in the current conservation equation~\eqref{app:eq:current101} and find the equations for $\hat{g}_1$,
\begin{align}
    \hat{\sigma}_z&= \hat{K}_1 + \sum_{i = 2}^{N} |m_{j1}|^2 \hat{U}_{j1}^{\dagger} \hat{K}_1 \hat{U}_{j1},
    \label{app:eq:current}\\
    \hat{K}_1 &= \hat{g}_1^{\dagger}\hat{\sigma}_z\hat{g}_1.
    \label{app:eq:K}
\end{align}
We decompose $\hat{K}_1 = \sum_{i=1}^{3} w_i \hat{\sigma}_i$, so Eq.~\eqref{app:eq:current} is linear in $(w_1,w_2,w_3)$, and determines a unique set of coefficients which we substitute in Eq.~\eqref{app:eq:K} to compute the solution
\begin{align}
    \hat{g}_{1} = \frac{\sqrt{w_z + \sqrt{w_x^2 + w_y^2 + w_z^2}}}{\sqrt{2}} \left( \hat{\mathbbm{1}} + i \frac{w_z - \sqrt{w_x^2 + w_y^2 + w_z^2}}{w_x^2 +w_y^2}(w_y \hat{\sigma}_x - w_x \hat{\sigma}_y )\right).
\end{align}
In this procedure there is a freedom in the choice of the $m_{j1}$ coefficients in \eqref{app:eq:current} and in Fig.~\ref{fig:transmittance}(d) we consider that $m_{j1} = 1$ if a state can scatter to cone $j$ and zero otherwise.

\subsection{Scattering from a normal electron gas to a Weyl semimetal}
\label{app:3deg}

\begin{figure}[htb]
    \centering
    \includegraphics[width=0.5\textwidth]{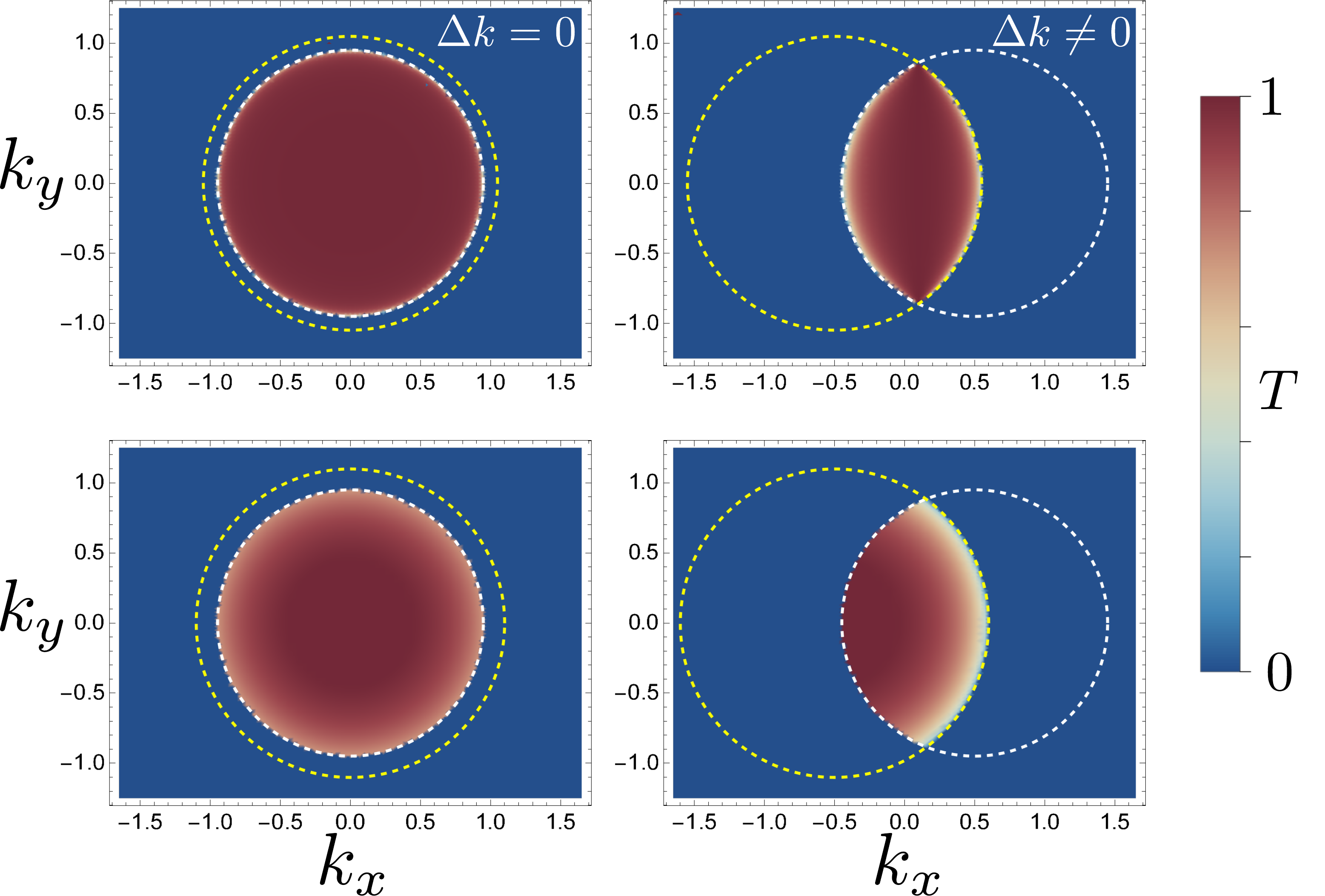}
   \caption{Transmittance $T$ at the Fermi energy, as a function of in-plane momenta ($k_x,k_y$) from a non-relativistic electron gas to another (first row) and from a non-relativistic electron gaz to a single ($N=1$) Weyl node (second row). In the figures, the positions of the Fermi surface for $z < 0$ and $z > 0$ are respectively $\epsilon_{0,L} = 0.9$ and $\epsilon_{0,R} = 1.1$.}
   \label{app:fig:transmittance}
\end{figure}

Here we consider the scattering from a non-relativistic electron gas, with Hamiltonian $\hat{H}_{\rm NR} = (k_x^2+k_y^2+k_z^2)/2m$, to a Weyl semimetal with $N$ Weyl nodes, described by independent wavefunctions $\phi_i$, $i \in [1,N]$. The current conservation now reads, instead of Eq.~\eqref{app:eq:current101},
\begin{align}
    \langle \Psi | \frac{1}{2mi}\left( \overleftarrow{\partial}_z - \overrightarrow{\partial}_z \right)| \Psi \rangle = \sum_{i = 1}^{N} v_{z,i} \langle \psi_i | \hat{\sigma}_z |\psi_i \rangle.
\end{align}
where the wavefunctions are evaluated at $z = 0$. We map this conservation relation to a linear transformation writing
\begin{align}
    \psi_i = \left( 
        \begin{array}{c}
            g_{1i}\\
            g_{2i}
        \end{array}
    \right) \Psi.
    \label{eq:matchmetal}
\end{align}
where $g_{1i}$ and $g_{2i}$ are unknown coefficients that relate the scalar wavefunction in the metal $\Psi$ to the spinor $\psi_i$ in the Weyl semimetal. This expression implies a relationship between the wavefunction of each Weyl node of the form $\psi_{i} = \hat{M}_{i1} \psi_{1}$, with $\hat{M}_{i1} = m_{i1} \hat{U}_{i1}$ where $m_{i1}$ is a scalar and $\hat{U}_{i1}$ a unitary operator defined in Eq.~\eqref{app:eq:unitary}. The conservation of probability current then reads
\begin{align}
    \frac{1}{i}\left( \overleftarrow{\partial}_z - \overrightarrow{\partial}_z \right) &= g_1^{\dagger} \underbrace{2m\left( v_{z,1}\hat{\sigma}_z + \sum_{i = 2}^{N} v_{z,i} |m_{i1}|^2 \hat{U}_{i1}^{\dagger}\hat{\sigma}_z\hat{U}_{i1} \right)}_{\equiv \lambda \hat{P}^{\dagger} \hat{\sigma}_z \hat{P} } g_1\\
    &= \lambda\left( (\hat{P}g_1)_+^{\dagger}(\hat{P}g_1)_+ - (\hat{P}g_1)_-^{\dagger}(\hat{P}g_1)_- \right)
    \label{eq:projection}
\end{align}
where the underbraced operator is decomposed in a difference of positively defined operators by a diagonalization. These operators are projectors on the states $(\hat{P} g_1)^{\dagger}_{\pm} \equiv g_1^{\dagger}P_{\pm}$ where ${P}_{\pm}$ are the eigenvectors of eigenvalue $\pm \lambda$ for the underbraced operator. This way, projecting this equation on the basis of plane waves, we find
\begin{align}
    (\hat{P}g_1)_+ &= \Theta(k_z) \sqrt{\frac{|k_z|}{\lambda}} |{\bf k}\rangle\langle {\bf k}|,\\
    (\hat{P}g_1)_- &= \Theta(-k_z) \sqrt{\frac{|k_z|}{\lambda}} |{\bf k}\rangle\langle {\bf k}|,
\end{align}
so the equation for the scattering coefficients is
\begin{align}
    &g_1\cdot\left( \Psi_i + r \Psi_r \right) = t\psi_{t1}\\
    \implies& (\hat{P}g_1)\cdot\left( \Psi_i + r \Psi_r \right) = t \hat{P}\psi_{t1}\\
    \implies& \left( 
        \begin{array}{c}
            \theta(k_z^{\rm (i)}) \sqrt{|k_z^{\rm (i)}|/\lambda}\\
            \theta(-k_z^{\rm (r)}) \sqrt{|k_z^{\rm (r)}|/\lambda}~ r \Psi_r
        \end{array}
    \right)
    = t \left(
        \begin{array}{c}
            (\hat{P}\psi_{t1})_+\\
            (\hat{P}\psi_{t1})-
        \end{array}
    \right)\\
    \implies& r = \sqrt{\frac{|k_{z}^{\rm(i)}|}{|k_z^{\rm (r)}|}} \frac{(\hat{P}\psi_{t1})_-}{(\hat{P}\psi_{t1})_+}.
\end{align}
From this expression we compute the reflectance $R =|r|^2$ and deduce the transmittance $T = 1 - R$. For example, for the interface with a single Weyl node, $\hat{P} = \hat{\mathbbm{1}}$ and $\lambda = v_{z,1}$ in Eq.~\eqref{eq:projection}, so we obtain
\begin{align}
    r = \frac{\beta({\bf k})}{\alpha({\bf k})} = \sqrt{\frac{1-v_z k_{z,t}/(E-\epsilon_{0,R})}{1+v_z k_{z,t}/(E-\epsilon_{0,R})}} e^{i\phi},
\end{align}
where $\alpha$, $\beta$, $k_{z,t}$ and $\phi$ are defined in Eqs.~(\ref{eq:alpha}-\ref{eq:snell}) of the main text. Also, in this situation, the operators $g_{1}$ and $g_{2}$ in the matching condition~\eqref{eq:matchmetal} are
\begin{align}
    g_1 &= \Theta(k_z) \sqrt{\frac{k_z}{\lambda}} |{\bf k}\rangle\langle {\bf k}|,\\
    g_2 &= \Theta(-k_z) \sqrt{\frac{|k_z|}{\lambda}} |{\bf k}\rangle\langle {\bf k}|.
\end{align}

\section{Extraction of Weyl point parameters}
\label{app:abinitio}

We identify possible Weyl semimetal candidate materials starting from the list of compounds reported in Ref.~\cite{xu2020comprehensive}. There, all non-magnetic compounds reported in the ICSD~\cite{Hellenbrandt2004} are investigated for the presence of Weyl points. As the analyzed heterostructures have to be experimentally grown as thin films, we restrict the selection to binary compounds because they are more likely to be grown as pure and ordered films. 

We seek to determine several parameters for each Weyl point: their location in the Brillouin zone, their velocities, their energetic position, and chirality. 

For the calculations, we apply a hierarchy of methods starting with density-functional theory (DFT) calculations as implemented in the VASP package~\cite{vasp} using the generalized-gradient approximation~\cite{perdew_generalized_1996} to describe the exchange-correlation potential. In the next step, we construct maximally-localized Wannier functions (MLWFs) and the corresponding tight-binding Hamiltonian with the help of the Wannier90~\cite{Mostofi2008} package. As the Wannierization process can be complex to control, we applied an automated procedure, which varies the necessary parameters until a the Hamiltonian is accurate enough. Here, accuracy is defined by taking the energy difference between the DFT and the Wannier tight-binding band structures expressed as $E_n^{DFT}$ and $E_n^{MLWF}$, respectively. The comparison was done for an energy window of 2 eV around the Fermi energy. In the last step, we use the resulting Hamiltonian to find the location of band crossings with an implementation of the Nelder-Mead algorithm~\cite{nelder1965simplex}. At each crossing, the Berry curvature in its close vicinity is investigated for all cardinal directions, where the orientation, i.e. pointing inwards or outwards, gives the chirality of the Weyl point. For each identified Weyl point all necessary parameters are then evaluated. We also investigate the presence of trivial bands at the Fermi level.

As the exact position and the slopes of the Weyl cones are very sensitive to the numerical methods, a cross-check is performed using another DFT code, FPLO~\cite{Koepernik1999}. Here, we employ the Hamiltonians from Ref.~\cite{xu2020comprehensive} to evaluate the robustness of the calculated parameters. The full workflow is shown in Fig.~\ref{app:fig:workflow}.

\begin{figure}[htb]
\centering
\includegraphics[width=0.75\textwidth]{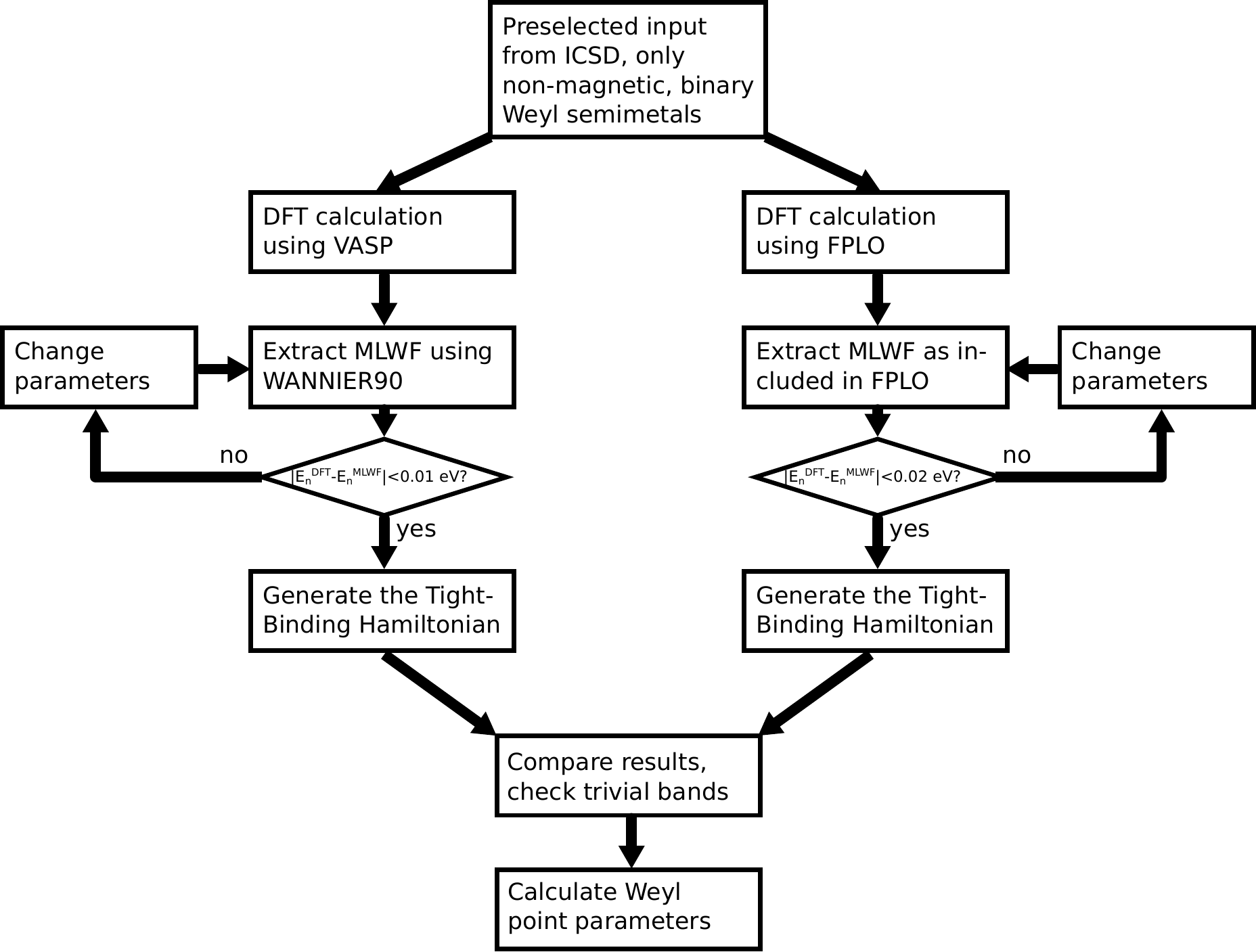}
   \caption{Workflow for the analysis of the Weyl points. The parameters to control the MLWF generation include the orbital projection and the inner and outer energy window for the disentanglement. The accuracy is calculated by comparing the energy eigenvalues of DFT, $E_n^{DFT}$, and of the Wannier interpolation, $E_n^{MLWF}$ in a window of 2 eV around the Fermi level. }
\label{app:fig:workflow}
\end{figure}
\end{widetext}
\end{document}